# MEASUREMENTS AND MODELISATION OF SHADOWING CROSS-CORRELATIONS BETWEEN TWO BASE-STATIONS

Karim ZAYANA, Bertrand GUISNET

France Télécom - Centre National d'Etudes des Télécommunications
6, avenue des usines BP 382, 90007 Belfort cedex, France

## 1. Introduction

Fading in mobile radio systems may be divided into two different types, fast and slow fading. Fast fading comes from multipath propagation whereas slow fading is caused by physical path loss and obstacles between the mobile and the base-station (shadowing). According to the Okumura model [1], the path loss attenuation is function of the distance logarithm of the propagation path. The slow fading expression (in dB) becomes:

$$A_{slow} = a + b \log d + A_{sh} \quad (1)$$

Parameters $a$ and $b$ depend on the environment. In urban area, typical values are $a = 16$ and $b = 36$ at 900 MHz [2], $d$ in meters.

Shadowing can be approximated by a log-normal law. Moreover, in [3], the author proposes a model taking into account autocorrelation proprieties of shadowing. This model has been adopted for radio-mobile systems testbed [4]. However, this model supposes that shadowing between a mobile and different base-stations are statistically independent. This assumption has not been checked yet. In the design of macroscopic diversity and hand-over schemes, and for a better understanding of the signal to interference distribution, the cross-correlation of the shadowing processes affecting different links should be studied too [5] [6].

Therefore, measurements in macro-diversity situations have been performed. Cross-correlation of shadowing terms between two base-stations and a mobile have been computed. It results that high cross-correlation coefficients may appear in practice. Afterwards, we propose a simple shadowing model, integrating both its autocorrelation and cross-correlation properties. Eventually, this model is used to compute simulations in some typical network cases.

## 2. Experimental configuration

### 2.1. Transmitters equipment

Macroscopic diversity measurements have been performed at 900 MHz in a small cell urban environment, in the French city of Mulhouse. For each experiment, two base-stations equipped with omni directional antennas (λ/2 dipoles) were set up. The two base-stations were approximately 700 m far. This is a typical distance between adjacent base-stations in a dense radio-cellular network. Antennas were about 30 m above ground level, and 10 m over the skyline of surrounding buildings. A carrier frequency was transmitted by each base-stations in the 900 MHz band. To distinguish the two signals, the frequency of one transmitter was 10 kHz shifted. The power supplied to each antenna was recorded in order to correct hypothetic variations. The transmitted power was set to 45 dBm. Two different base-station couples were probed ensuring a better robustness of the results.

### 2.2. Measurement proceeding

The receiver was placed in a vehicle, a $\lambda/4$ antenna fixed on its roof. Signals from both base-stations were measured simultaneously. The received signal was sampled every $\lambda/2$ (15 cm) during a run and filtered to extract the incoming information from each link. By using this method, the vehicle had not to be necessarily driven at constant speed to recover the exact position of the measurement points.

The 10 kHz difference of frequency has no impact on slow fading, which does affect uniformly the whole frequency bandwidth. It has no impact on fast fading either, as it is much smaller than the 90 % channel coherence bandwidth which is close to the MHz.

Measurement areas were selected in order to receive mean signal strengths sufficiently high (above -100 dBm, which allows path loss up to 145 dB). The measurements were performed in the surrounding area, up to 3 km far from the base-stations. More than 60 km of macroscopic diversity measurements were recorded.

## 3. Shadowing calculation

The shadowing contribution to the global attenuation had to be extracted. Fast fading was easily discarded by computing local means of the received field using a 40 $\lambda$ (12 m) sliding window. Then, we had to deduce the shadow fading from the slow fading. As expected and due to the large distance between transmitters, fast variations of the signals were found to be uncorrelated on the two branches [7].

We have tested several methods to extract shadowing :
- We first derived the propagation parameters *a* and *b* of equation (1) using a linear regression, the difference between the slow fading and the path loss being the shadow fading. Nevertheless, parameters *a* and *b* may vary from one place to another, and may depend of the base-station. So, the measured area was divided into four homogeneous parts. For each zone and each base-station, parameters *a* and *b* have been computed. This method fits to the initial definition of the shadowing, but it presents different and discontinuous results from one quarter to another. The resulting shadow fading was centered and had a 7 dB standard deviation. This value is lower than the one generally given in the literature (10 dB), because it was deduced from local measurements performed in a relatively small area (10 km$^2$ or so).

- We also computed the shadow fading by meaning the slow fading with a 800 m sliding window and by subtracting it to the original slow fading. Consequently, the shadow fading was derived from the fields measured in a same quarter, and often in a same street. This technique seems to take better account of the continuous variations in time of the shadow fading and of its physical interpretation. The resulting shadow fading is centered and presents a lower standard deviation ( 5,5 dB).

## 4. Measurement Analysis

### *4.1. Parameters of interest*

The cross-correlation factor has been computed on the whole data. It appears a low but non zero correlation. But, in some areas, high correlation coefficients were observed. Based on physical considerations, we studied the incidence of two geometrical parameters on the cross-correlation coefficients : the angle $\theta$ from which the vehicle sees the two base-stations, and the ratio between the distance from the mobile to each station, $R_{dB} = 10|\log(d_1/d_2)|$, figure 1. One can notice that the two parameters describe all the plane and form a system of coordinate.

We expect an increasing of the correlation factor as the portion of propagation paths in common increases, thus, as the angle $\theta$ decreases and, meanwhile, as the ratio $R_{dB}$ decreases too. Indeed, when the transmitter antenna is over roof-top, the base-station to mobile profile is the principal propagation path. For low values of $\theta$, and $R_{dB}$, an obstacle masking one base-station will probably obstruct the other one.

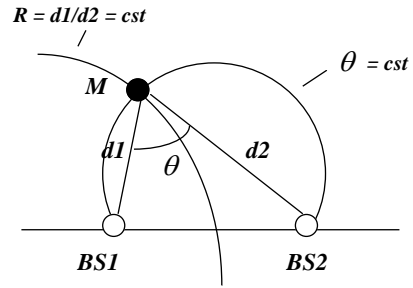

**Figure 1 : definition of $\theta$ et $R_{dB}$**

In a design of macro-diversity, simultaneous links may occur when path losses on the two branches are close, thus, when the parameter $R_{dB}$ is low (distances from the base-stations are of the same order). Macro-diversity is used to mitigate the effect of shadow fading. Its efficiency is maximum when shadow fading terms present a total decorrelation on the different branches. In literature, macro-diversity gain is evaluated assuming total independent of shadowing terms [5], which could overestimate performances of macroscopic diversity systems.

The dependency of the correlation, function of $\theta$ and $R_{dB}$, may also be interesting to study interferences caused by one station to another. In the hypothesis of high shadowing cross-correlation, the useful and interfering signals are varying in the same way, so that the range of the C/I remains small compared to the uncorrelated case.

*4.2. Shadow fading analysis*

We processed tables of cross-correlation coefficients versus parameter $\theta$, table 1. Few differences were observed according to the way shadow fading was computed. As expected, in average, top values of cross-correlation (over 0.7) have been found when the mobile is in the axis of the two base-stations, or stays slightly aside with and angle between 0 and 30°. On the other hand, low values (below 0.2) were found when the mobile is located between the two stations.

Looking more accurately, we noticed that streets orientation could be taken into account. Actually, correlations seem higher when streets are parallel to the base-stations axis, and lower when orthogonal to this same axis.

Additional measurements with macroscopic diversity have just been processed in Paris and should help us to sharpen this first analysis.

Simulations were also achieved with a COST 231 derived model developed by the CNET (France Télécom) [8] on PARCELL simulator ( PARCELL is France Télécom radio cellular engineering tool ). This provided an important pool of statistical samples predicted in Paris or Mulhouse, which allowed us to get more precise cross-correlation tables involving both parameters $\theta$ and $R_{dB}$, table 2.

| | $\theta \in [0° ;30°]$ | $\theta \in [30° ;60°]$ | $\theta \geq 90°$ |
|---|---|---|---|
| | $\alpha=0,6$ | $\alpha=0,25$ | $\alpha \geq 0.2$ |

**Table 1 : measured cross-correlations**

| | $\theta \in [0° ;30°]$ | $\theta \in [30° ;60°]$ | $\theta \in [60° ;90°]$ | $\theta \geq 90°$ |
|---|---|---|---|---|
| $R_{dB} \in [0 ;2]$ | $\alpha=0,8$ | $\alpha=0,5$ | $\alpha=0,4$ | $\alpha=0,2$ |
| $R_{dB} \in [2 ;4]$ | $\alpha=0,6$ | $\alpha=0,4$ | $\alpha=0,4$ | $\alpha=0,2$ |
| $R_{dB} \geq 4$ | $\alpha=0,4$ | $\alpha=0,2$ | $\alpha=0,2$ | $\alpha=0,2$ |

**Table 2 : Predicted cross-correlations**

However, the shadow fading standard deviation resulted much lower in simulations than in practice (4 dB instead of 5.5 or 7 dB). Indeed, even if the propagation model present very good performances, it does not reproduce all the shadowing variations.

Despite these differences, this simulation confirms results obtained on measurements and their reliability.

## 5. Shadow fading model and applications

*5.1. Shadow fading modelisation*

We propose a model of the shadow fading, including both its autocorrelation and cross-correlation properties. This model aims to be implemented in system and capacity simulators for example.

Let one mobile and $N$ surrounding base-stations, considered either as diversity or interfering transmitters. We can divide the map into large areas where it is assumed that correlation coefficients are more or less constant. Cross-correlation coefficients are given by our previous analysis. Of course, autocorrelation coefficients can be found from our own measurements, but this subject of interest has already been studied in the literature, [3]. Indeed, it is generally assumed that the shadow fading process (in dB) is gaussian and centered, and that its autocorrelation decreases exponentially as the distance between measurement points increases.

We call $s_i(kT)$ the shadow fading value affecting channel number *i* at instant *kT*, with $1 \leq i \leq N$ and $k \in \mathbf{N}$. We wish to generate $N$ pseudo random shadowing processes $s_i$, $1 \leq i \leq N$, satisfying the following correlation properties :

$$\langle s_i(kT) s_i(lT) \rangle = \sigma^2 \beta^{|l-k|} \qquad (2)$$
(auto-correlation property)

and :

$$\langle s_i(kT) s_j(kT) \rangle = \sigma^2 \alpha_{i,j} \qquad (3)$$
(cross-correlation property)

where $\sigma^2$ is the shadow fading variance, $\beta$ the autocorrelation coefficient between two consecutive shadow fading terms, and $\alpha_{i,j}$, $1\leq i\leq N$, $1\leq j\leq N$, the cross-correlation coefficients between the different links.

We first produce $N$ independent processes with good autocorrelation properties by filtering $N$ independent white gaussian noise $g_i$, $i \in [1;N]$, of variance $\sigma^2$. We have for each $i$ and all $k$ :

$$b_i(kT) = \beta\, b_i((k-1)T) + \sqrt{1-\beta^2}\, g_i(kT) \quad (4)$$

The covariance matrix $\mathbf{M} = (\alpha_{p,q})_{\substack{1\leq p\leq N \\ 1\leq q\leq N}}$ is known to be symmetrical and positive. Therefore, it can be reduced into the Cholesky form $\mathbf{M} = \mathbf{H}\,^t\mathbf{H}$ where $\mathbf{H}$ is triangular [9]. Let $\mathbf{B}(kT)$ the column-vector of processes $b_i$, $i \in [1;N]$, at instant $kT$. Then, we compute the product $\mathbf{S}(kT) = \mathbf{H}.\mathbf{B}(kT)$. It is easy to cheek that :

$$\langle \mathbf{B}(kT)\,^t\mathbf{B}(lT) \rangle = \beta^{|l-k|} \mathbf{I}_N$$

so that :

$$\langle \mathbf{S}(kT)\,^t\mathbf{S}(lT) \rangle = \sigma^2 \beta^{|l-k|} \mathbf{M}$$

Calling $s_i(kT)$, $i \in [1;N]$, the components of $\mathbf{S}$ at instant $kT$, we deduce :

$$\langle s_i(kT) s_j(lT) \rangle = \sigma^2 \alpha_{i,j} \beta^{|l-k|} \quad (5)$$

Relations (2) and (3) are then satisfied.

### 5.2. Simulations of C/I

We have used our shadow fading model to investigate statistical properties of the C/I parameters. We have located on a map one base-station and several co-channel interfering base-stations in some typical and realistic cases. Thanks to Monte Carlo simulations, we could compute the mean and the standard deviation of the C/I on some precise places. Parameters $a$ and $b$ of the path loss attenuation were fixed to the usual values $a = 16$ and $b = 36$.

Notice that the decorrelation distance of the shadow fading has no impact on the results of the simulations achieved. But for more complex studies, involving power control for example, it should be taken into account.

An example of configuration with two jamming stations N1 and N2 and one source S is given on figures 2 and 3. In the first case, no correlation is introduced, whereas in the second one, cross-correlation coefficients are read from table 2. The standard deviation of the shadow fading is 7 dB.

By performing various simulations, we noticed that an increase of the cross-correlation factors always provoked an increase of the mean of the C/I (between 0 and 1 dB) and a decrease of its standard deviation (between 0 and 2 dB). Thus, predicting mobile system performances without considering shadow fading correlations could lead to pessimistic results.

We also remarked that an increase of the number of jamming stations provoked a decrease of the mean of course, but also a slight decrease of the standard deviation.

Eventually, taking a higher standard deviation parameter $\sigma$, for instance 10 dB instead of 7 dB, causes a lower mean of the C/I (-1 dB sometimes), and an higher standard deviation. Then, predictions of quality of a mobile system would be affected by this change.

| m=9,8 σ=8,8 | m=8,2 σ=8,8 | m=3,2 σ=8,9 | m=-3,7 σ=8,9 | N2 |
| m=17,2 σ=9,0 | m=18,3 σ=9,0 | m=9,8 σ=8,9 | m=-0,6 σ=9,0 | |
| m=24,9 σ=8,9 | m=49,3 σ=8,9 | m=15,3 σ=8,9 | m=0,6 σ=9,3 | |
| | | S | | N1 |

**Figure 2 : Statistics of C/I, uncorrelated case**

|         |         |         |         | N2 |
|---------|---------|---------|---------|----|
| m=10,7  | m=9,1   | m=3,9   | m=-3,1  |    |
| σ=8,2   | σ=8,2   | σ=8,1   | σ=7,4   |    |
| m=18,1  | m=19,1  | m=10,4  | m=0,0   |    |
| σ=8,3   | σ=8,4   | σ=8,0   | σ=7,8   |    |
| m=25,5  | m=49,9  | m=15,7  | m=0,8   |    |
| σ=8,2   | σ=8,2   | σ=8,1   | σ=7,9   |    |
|         | 🟢 S    |         |         | N1 |

**Figure 3 : Statistics of C/I, correlated case**

## 6. Conclusion

Propagation measurements and simulations devoted to the shadow fading analysis in a diversity context have been presented. This work gives a more accurate understanding of the cross-correlation phenomena. It will be completed shortly by taking into account new measurements perform in Paris. It already emerges that cross-correlation coefficients may be quite high when the two stations see the mobile with the same azimuthal angle. In return, a complete decorrelation is often guaranteed when the mobile is between the two stations.

The tables derived can be used either to pursue studies about macro-diversity or to investigate further on the C/I statistical properties. The realistic modelisation proposed above is also particularly interesting to evaluate performances of hand-over algorithms and to estimate macro-diversity gain. For both cases, we presented a model of the shadow fading processes, taking into account their different correlation characteristics. We applied this model to perform some simulations.